\documentclass[aip, amsmath,amssymb, reprint]{revtex4-1}
\usepackage{graphicx}
\usepackage{amsmath,braket,mathdots}
\usepackage{dcolumn}
\usepackage{bm}
\usepackage[utf8]{inputenc}
\usepackage[T1]{fontenc}
\usepackage{mathptmx}
\usepackage{etoolbox}

\usepackage{subfigure}
\usepackage{xcolor}
\usepackage{float}

\makeatletter
\def\@email#1#2{%
 \endgroup
 \patchcmd{\titleblock@produce}
  {\frontmatter@RRAPformat}
  {\frontmatter@RRAPformat{\produce@RRAP{*#1\href{mailto:#2}{#2}}}\frontmatter@RRAPformat}
  {}{}
}%
\makeatother

\begin{document}

\title{
Bifurcations and Intermittency in Coupled Dissipative Kicked Rotors
}
\author{Jin Yan}
\affiliation{\small Weierstrass Institute for Applied Analysis and Stochastics, Mohrenstra{\ss}e 39, 10117 Berlin, Germany}
\email{jin.yan@wias-berlin.de}

\begin{abstract}
We investigate the emergence of complex dynamics in a system of coupled dissipative kicked rotors and show that critical transitions can be understood via bifurcations of simple states. We study multistability and bifurcations in the single-rotor model, demonstrating how these give rise to a variety of coexisting spatial patterns in a coupled system. A combined order parameter is introduced to characterize different spatial patterns and to reveal the coexistence of chaotic and regular attractors. Finally, we illustrate an intermittent phenomenon near the onset of chaos. 
\end{abstract}

\maketitle

{\bf 
Phase transitions in coupled dynamical systems are intricate due to high dimensionality, nonlinearity and multistability. Understanding critical transitions is nontrivial and a general approach involves employing statistical mechanical (or macroscopic) quantities. 
In this study we offer a microscopic perspective. 
We begin with a systematic study of dynamics of the single-rotor model, emphasizing bifurcations and basins of attraction of multiple attractors. These results significantly contribute to understanding coexisting spatiotemporal patterns in the coupled system. The bifurcations of simple states establish critical boundaries where more complex spatial patterns emerge, characterized using a combined order parameter adapted from Kuramoto phase oscillator models. Lastly, we illustrate an intermittent behavior observed near the onset of chaos, which belongs to type-I super-transient (i.e., chaos spreads out in a percolation-like manner). 
Our findings highlight that, despite the multistability in the coupled system, transitions in the dominant (or physically observable) behavior can be well-captured by bifurcations of simple states, along with the dynamics at the single-element level. 
}

\section{Introduction} 
\label{sec-intro}
Coupled dynamical systems serve as a fundamental framework for understanding a wide range of complex phenomena in physics, biology and engineering. These systems, characterized by interacting components, exhibit intricate behaviors that emerge from interplay between individual dynamics and the coupling mechanisms. 

Multistability, where a system can have multiple stable states under the same set of parameter values, is prevalent in coupled systems. 
In natural systems such as ecosystems, the ability of a system to shift between stable states (for example, forest and grassland) has profound implications for biodiversity and resilience \cite{mitra2015integrative}. Climate systems exhibit multistability with distinct states like ice ages and warm interglacial periods \cite{feudel2018multistability, margazoglou2021dynamical}. In neuroscience, memory and decision-making often rely on multistable patterns in neural circuits \cite{kelso2012multistability}. Insights are given in control systems to avoid unintended state shifts such as failures in power grid \cite{kim2018multistability, delabays2022multistability}, chaos in Josephson junction arrays \cite{louodop2019extreme, ramakrishnan2021suppressing} and turbulence in plasma flows \cite{teaca2023overview}. 
In social sciences, multistability can model phenomena like cultural shifts, economic cycles and political polarization \cite{cavalli2016complex,ferraz2023multistability}. 
Recognizing when a system is near a critical point can help predict or prevent undesirable outcomes. 

In many cases, the coexistence of multiple stable states makes the system sensitive to perturbations or parameter variations, and give rise to changes in basins of attraction \cite{shrimali2008nature,feudel2008complex}. 
For example, crisis bifurcations can occur when a chaotic attractor in a multistable system collides with a basin boundary, leading to abrupt changes in the system behavior \cite{grebogi1986critical,grebogi1987critical}.
Therefore, bifurcations act as mechanisms that create, modify or eliminate multistable states. Even a simple one-dimensional dynamical system can have a saddle-node (or fold) bifurcation that generates or annihilates a pair of stable and unstable states, changing the number of coexisting attractors \cite{kuznetsov1998elements}. 
In large interacting dynamical systems, understanding how bifurcations in the single element translate into phase transitions in the collective behavior of the coupled system is challenging. 

Coupled map lattices, one of the simplest mathematical models for spatially extended systems, where the continuous dynamical variables are on the discrete (lattice) space with discrete time, have been studied extensively \cite{kaneko1989pattern, kaneko1989diffusion, oliveira2014statistical, yan2024prethermalization}. They are a powerful framework for modeling and understanding spatiotemporal complexity in high-dimensional systems. They combine simple local nonlinear dynamics with spatial coupling, making them ideal for exploring how interactions lead to emergent collective behavior. However, while coupled map lattices are often analyzed through macroscopic observables (e.g., global order parameters and Lyapunov spectra), a comprehensive microscopic understanding remains elusive.  

A recent study \cite{russomanno2023spatiotemporally} identified interesting phase transitions in a system of locally coupled dissipative kicked rotors by exploring statistical observables such as variance of the momentum distribution, averaged kinetic energy and largest Lyapunov exponents. Their phase diagram indicated transitions among ``trivial", ``pattern", ``spatiotemporal ordering" and chaotic states, yet the underlying mechanisms remain unclear, particularly in relation to the bifurcations in the single-rotor model. 

\begin{figure*} 
\includegraphics[width=0.95\linewidth]{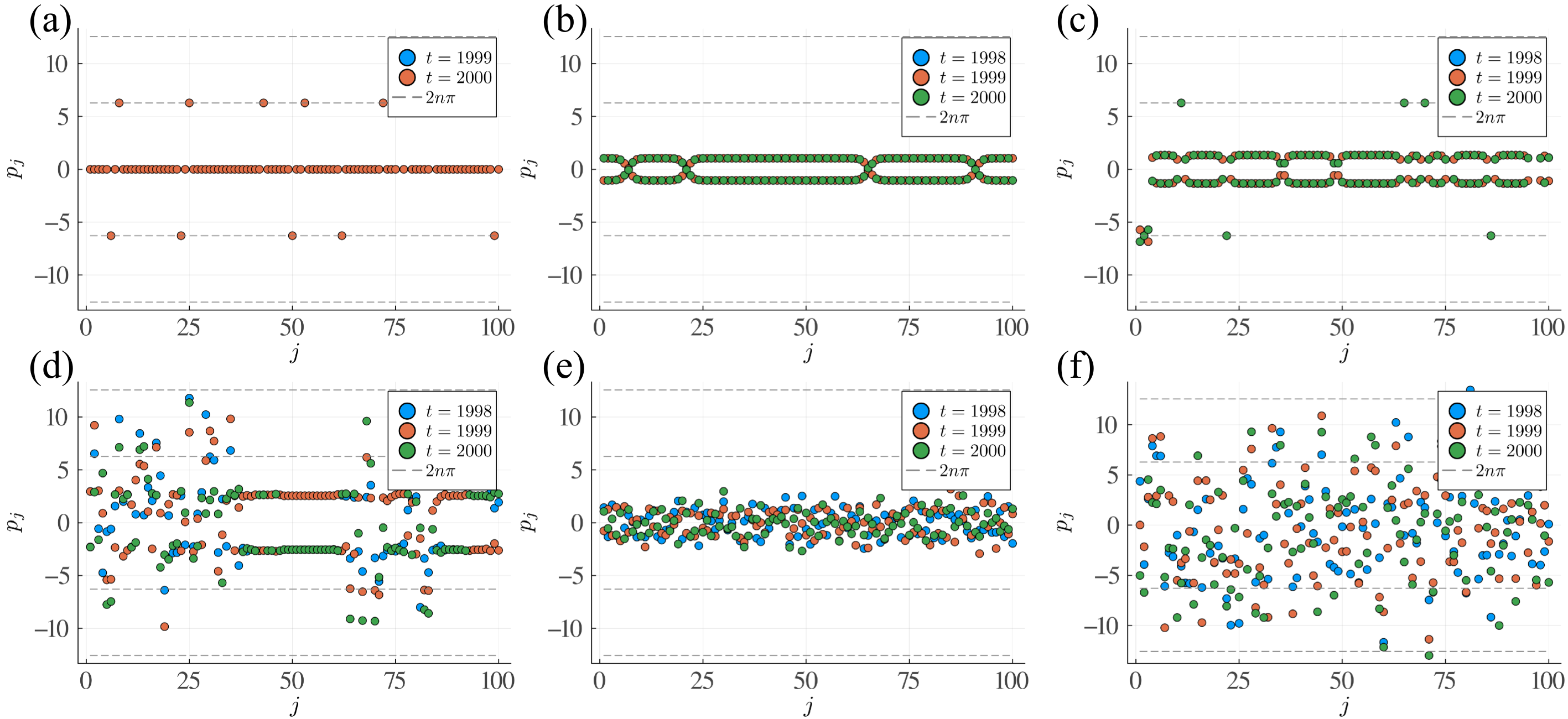}
\caption{\label{fig-intro} Distinct typical snapshots in momenta $\{p_j\}$ ($j = 1, 2, ..., 100$) of the coupled dissipative kicked rotor system Eq.\eqref{eq-ckr}, each drawn from a random initial condition with parameter values (a) $K_0=1.9, J=0.3$, (b) $K_0=1, J=0.8$, (c) $K_0=2, J=0.6$, (d) $K_0=4.8, J=0.2$, (e) $K_0=0.6, J=1.1$, (f) $K_0=5, J=0.5$. }
\end{figure*}

Simulations drawn from random initial conditions already indicate that ``pattern" and ``spatiotemporal ordering" have different aspects. In Fig.\ref{fig-intro}, we illustrate six different snapshots of the rotor momenta, corresponding to different parameter regions. 
The patterns on the first row clearly exhibit spatial and temporal periodicity, whereas those on the second row appear chaotic in time, space or both. 
Many questions arise: how does the stationary (or frozen) state observed in Fig.\ref{fig-intro}(a) transition into temporal period-$2$ states featuring spatially alternating patterns, as illustrated in Figs.\ref{fig-intro}(b)-(c)? How are these transitions connected to the dynamics of the single-rotor model? And how can these patterns be effectively captured both spatially and temporally? 

These questions motivate us to investigate microscopic dynamics and analyze bifurcations in the simplest states, as more complex states emerging from random initial conditions can be understood through these elementary states and the multistability inherent in the single-rotor model. 
Our advancements can be summarized as follows: 

(i) the detailed bifurcations in the single dissipative kicked rotor is analyzed, which provides key insights into the patterns observed in the weakly coupled system; the coexistence of the chaotic and regular attractors already exists at the level of a single rotor; 

(ii) even for small couplings, there exist multiple stable states depending on the nonlinearity parameter and initial conditions (an example is shown in Fig.\ref{fig-intro}(a)); however, if all initial momenta are restricted in a small interval near zero, a unique homogeneous-zero state is expected; 

(iii) the homogeneous-zero state bifurcates into a spatial period-$2$ and temporal period-$2$ state, which we will refer as the alternating state; within the stability region of the alternating state, multiple periodic patterns emerge (examples are shown in Figs.\ref{fig-intro}(b)-(c)) and can be captured by an order parameter;

(iv) the coexistence of chaotic and regular states is also observed in the coupled system near the onset of chaos, exhibiting spatiotemporal intermittency and type-I super-transient, where chaos spreads out in a percolation-like manner \cite{crutchfield1988attractors}.

The numerical method used in this study for detecting bifurcations is pseudo arclenth continuation (PALC), encoded in the Julia package BifurcationKit.jl \cite{veltz:hal-02902346}. 

The paper is organized as follows. We first introduce in detail the model of a single rotor in Sec.\ref{sec-sr}, including the cascades of bifurcating branches, their basins of attraction and probability distributions on chaotic attractors. In Sec.\ref{sec-ckr} of the coupled rotor system, we first show stability of the simplest possible state (i.e., homogeneous-zero state) and its bifurcated state (alternating state) in Sec.\ref{subsec-simple-regular}. Then in Sec.\ref{subsec-patch} we study less trivial spatial states and classify them as two different patched states. In Sec.\ref{subsec-order-parameter} we employ a combination of Kuramoto and Daido order parameters to characterize spatial symmetry before the transition to chaos. In Sec.\ref{subsec-intermittency} we illustrate long transient behavior with spatiotemporal intermittency and a transition to chaos. Finally in Sec.\ref{sec-conclu} we draw conclusions and give an outlook.

\section{Single dissipative kicked rotor}
\label{sec-sr}
To understand the complex spatiotemporal patterns generated by the coupled system in Fig.\ref{fig-intro}, we first study the single-rotor dynamics. 
A single kicked rotor with dissipation was first introduced by George M. Zaslavsky
\cite{zaslavsky1978simplest} which is now also called the {\it Zaslavsky map} (or dissipative standard map). It was derived from perturbing a stable limit cycle of an oscillator by an external periodic force. The stroboscopic map for the rotor angle $\theta \in [-\pi, \pi]$ and its angular momentum $p \in \mathbb{R}$ can be written as 
\begin{align}
p(t+1) &= \gamma p(t) - K_0\sin \theta(t), \label{eq-zaslavsky1}
\\
\theta(t+1) &= \theta(t) + p(t+1) \quad (\text{mod } 2\pi), 
\label{eq-zaslavsky2}
\end{align}
where $\gamma \in [0, 1]$ is the dissipation coefficient, $K_0 > 0$ is the nonlinearity parameter, and time $t \in \mathbb{N}_0$.

When $\gamma=0$, the system reduces to the one-dimensional Arnold circle map $\theta(t+1) = \theta(t) - K_0\sin \theta(t) \quad (\text{mod } 2\pi)$ -- a fundamental model of phase locking \cite{boyland1986bifurcations}. In contrast, for $\gamma = 1$ (no dissipation), it becomes the Chirikov standard map, a classical low-dimensional example of Hamiltonian chaos exhibiting KAM tori, stochastic layers and chaotic seas \cite{chirikov1979universal, lichtenberg2013regular}.

The Jacobian determinant equals to $\gamma$, such that for $\gamma>0$ the map is invertible and can be regarded as a Poincare map of some three-dimensional flow \cite{ivankov2001complex}. 
For $\gamma \in (0, 1)$ the Zaslavsky map has an attractor which for sufficiently large $K_0$ is known to be  chaotic \cite{zaslavsky1978simplest}, with contraction along $p$ (due to dissipation) and expansion along $\theta$. Such an attractor is shown on the first row of Fig.\ref{fig-chaos} in Appendix \ref{appA2}.

\subsection{Cascades of bifurcating branches}
\label{subsec-sr-bifur}
First, we study regular (non-chaotic) solutions of the Zaslavsky map. For an $m$-periodic solution we define the average momentum  
 \begin{equation*}
 \bar{p} := \frac{1}{m}\sum_{j=1}^m p(j).
 \end{equation*}
Note that Eq.\eqref{eq-zaslavsky2} implies that an $m$-periodic trajectory satisfies
$$ 
\theta(t+m) - \theta(t) = \sum_{j=1}^m p(t+j) \quad (\text{mod } 2\pi).
$$
Hence, for an $m$-periodic solution we can conclude that the product $m\bar{p}$ is an integer multiple of $2\pi$. We will use the numbers $(m,\bar{p})$ to characterize the periodic solutions. It turns out that they are organized in a cascade of branches where, for increasing $K_0$, each branch undergoes a period-doubling sequence, while  $\bar{p}$ remains fixed. There are main $n$-resonances, where $\bar{p}=2\pi n$ for some $n\in\mathbb{Z}$. They start with fixed points $m=1$ in a fold bifurcation. We show now how the folds as well as the first period doublings  can be calculated explicitly.

A fixed point $(p^*, \theta^*)$ of the Zaslavsky map satisfies
\begin{align*}
p^* &= 2n\pi,\\
K_0\sin \theta^* &= 2n\pi(\gamma-1).
\end{align*}
This allows to calculate the characteristic equation for multipliers $\lambda\in\mathbb{C}$ of the Jacobian at the fixed points as
$$\lambda^2 - (\gamma + 1 - K_0\cos \theta)\lambda + \gamma = 0.$$
Inserting $\lambda=\pm 1 $, we obtain the bifurcation conditions for the fold and the period-doubling (PD) as 
\begin{align*}
K_0^{\text{fold}, n} &= \pm 2n\pi(1 - \gamma),\\
K_0^{\text{PD}, n} &= \pm 2\sqrt{(1 + \gamma)^2 + (1 - \gamma)^2n^2\pi^2}.
\end{align*}
The corresponding bifurcating points are therefore given by 
\begin{equation*}
p(\gamma, K_0) = 2n\pi = \mp \frac{K_0}{1 - \gamma}, \quad \theta(\gamma, K_0) = \pm \frac{\pi}{2}, 
\end{equation*} 
for the fold, and 
\begin{equation*}
\begin{split}
p(\gamma, K_0) &= 2n\pi = \mp \frac{\sqrt{K_0^2 - 4(1 + \gamma)^2}}{1 - \gamma}, \\
\theta(\gamma, K_0) &= \pm \arccos \frac{2(1 + \gamma)}{K_0}, 
\end{split}
\end{equation*}
for the period-doubling. These branches are shown as dashed red and blue curves in Fig.\ref{fig-bifur} for a fixed $\gamma=0.8$ (see Appendix \ref{appA} for other $\gamma$ values). 
Notice that although the phase space is an infinite cylinder $(p, \theta) \in  \mathbb{R}\times [-\pi, \pi]$, the red radial lines bound the rotor momentum; this radial region is narrowed down when the dissipation is enhanced, i.e., when $\gamma$ decreases, cf. Appendix \ref{appA}. The discrete $p^*=2n\pi$ ($n \in \mathbb{Z}$) branches somewhat resemble discrete energy levels of an atom, and as $K_0$ increases, more admissible levels appear. When chaos appears, there is coexistence of main $n$-resonances and a (bounded) chaotic attractor. However, as $K_0$ increases the distance between the dashed blue and red curves decreases; there exist windows of $K_0$ for which the chaotic attractor is the only attractor in the system, see also Sec.\ref{subsec-sr-chaos}. 

Other branches of nontrivial periodic solutions  ($m \neq 1$) also emerge in a fold bifurcation and, for larger values of $K_0$ undergo period doublings. These bifurcations can be found by numerical bifurcation analysis based on the continuation method mentioned in Sec.\ref{sec-intro}. 

An example is shown in light-green in Fig.\ref{fig-bifur}, where a period-$3$ orbit (a subharmonic $(n, m)=(1, 3)$) undergoes period doublings. The dashed green curve obtained by the continuation method shows such a bifurcation scenario for higher $n$. Another example shown in cyan is a period-$4$ orbit on the main $0$-resonance, or a subharmonic $(n, m)=(0, 4)$. 

\begin{figure*}
\includegraphics[width=0.65\linewidth]{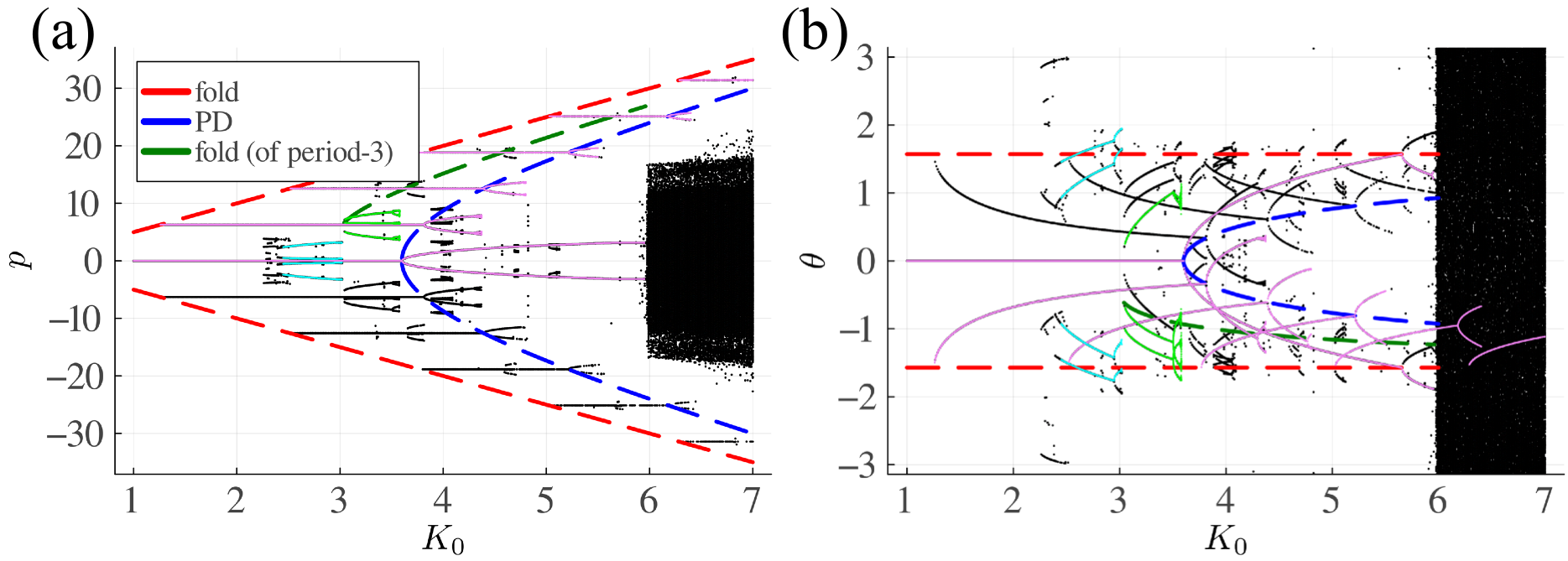}
\caption{\label{fig-bifur} Bifurcations in the Zaslavsky map with $\gamma = 0.8$ and $K_0 \in [1, 7]$: (a) for momentum $p$ and (b) for angle $\theta$. The dashed curves connect a cascade of bifurcation points: fold (in red) and period-doubling (PD in short, in blue) bifurcations for fixed points $p^* = 2n\pi$, $n \in \mathbb{Z}$, and fold bifurcations for a period-$3$ state (in dark-green). Each plot also highlights the main $n$-resonances in violet, a subharmonic $(n, m) = (0, 4)$ resonance in cyan, and a subharmonic $(n, m) = (1, 3)$ resonance in light-green. Bifurcations for other $\gamma$ values are presented in Appendix \ref{appA}.}
\end{figure*}

From now on, we fix $\gamma = 0.8$ for all numerical illustrations.

\subsection{Basins of attraction} 
\begin{figure*}
\includegraphics[width=0.99\linewidth]{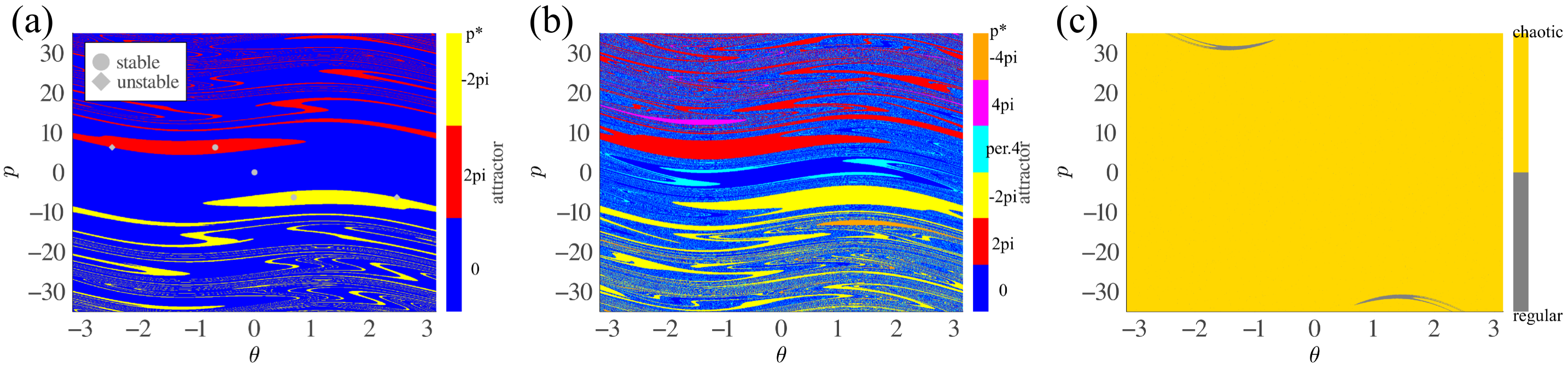} 
\caption{\label{fig-basins} Basins of attractions of the Zaslavsky map with $\gamma = 0.8$ and (a) $K_0 = 2$, (b) $K_0 = 2.7$ and (c) $K_0 = 6.6$.}
\end{figure*}

Fig.\ref{fig-basins} shows basins of attraction of various stable states in the Zaslavsky map. For $K_0 = 2$, the only attractors are the three fixed points $p^* = 0$ and $\pm 2\pi$ (in gray circles). The two unstable fixed points (in gray diamonds) are also highlighted in Fig.\ref{fig-basins}(a). 
The basin boundaries exhibit complicated and highly nonlinear features. 

For $K_0 = 2.7$, there appear other attractors whose basins are labeled in different colors in Fig.\ref{fig-basins}(b). 
Specifically, apart from the three colors presented in Fig.\ref{fig-basins}(a), we have basins for a period-$4$ orbit (in cyan, corresponds to the same color in Fig.\ref{fig-bifur}), for fixed point $p^* = 4\pi$ (in magenta) and for fixed point $p^* = -4\pi$ (in orange). 
In the chaotic regime ($K_0 = 6.6$), the basins for the regular branches $p^* = \pm 10\pi$ are very small, as shown in gray in Fig.\ref{fig-basins}(c). 

\subsection{Chaotic regime}
\label{subsec-sr-chaos}
We notice that chaos is emerged from successive bifurcations of the trivial fixed point $(p^*, \theta^*) = (0, 0)$. 
Interestingly, the chaotic attractor can coexist with a pair of regular branches that is bounded by the red and blue curves in the bifurcation diagram of $p$. For example, at $K_0=6.6$, the regular branches given by $p = 5\cdot (2\pi) \approx \pm 31$ coexist with the chaotic attractor, see Appendix \ref{appA2} for probability measures on these attractors. Such coexistence with the regular branches is no longer observed due to the decreasing distance between the dashed red and blue curves in Fig.\ref{fig-bifur}. 

As illustrated in \cite{russomanno2023spatiotemporally}, the map undergoes a period-doubling cascade to chaos, but the cascade occurs almost immediately and does not follow the Feigenbaum constant universality, unlike the logistic map or other one-dimensional unimodal maps \cite{strogatz2024nonlinear}.

In this section, we examined rich dynamics of the single-rotor model, including cascades of bifurcations and coexistence of regular and chaotic states. In the next section, we explore the coupling of these rotors, show how the dynamics at the single-rotor level is reflected and new patterns are created in the spatially extended systems, and therefore give explanations to the complex spatiotemporal patterns illustrated in Fig.\ref{fig-intro}.

\section{Coupled dissipative kicked rotors}
\label{sec-ckr}
We consider a system $\boldsymbol{F}$ of $N$ coupled identical dissipative kicked rotors, whose dynamics is given by 
\begin{equation}
\begin{split}
\begin{cases} 
p_j(t+1) = \gamma p_j(t) - K_0\sin \theta_j(t) + J\Delta_j(t) \\
\theta_j(t+1) = \theta_j(t) + p_j(t+1) \quad (\text{mod } 2\pi) 
\end{cases} \\
\Delta_j(t) := \sin \left(\theta_{j-1}(t) - \theta_j(t)\right) + \sin \left(\theta_{j+1}(t) - \theta_j(t)\right),
\end{split}
\label{eq-ckr}
\end{equation}
where the coupling $\Delta_j(t)$ is considered through the sine of differences between the nearest neighboring rotors at time $t \geq 0$; $j = 1, 2, ..., N$ labels the rotors with periodic boundary conditions, and $J \geq 0$ is the coupling strength. 
There is no physical reason that $J$ has to be non-negative, but we leave the negative $J$ case for future work. 

Some statistical properties of this coupled system have already been addressed in \cite{russomanno2023spatiotemporally}. In the following sections we focus on bifurcations of simple states, spatial patterns and spatiotemporal intermittency near the transition to chaos.

\subsection{Simple regular states and their bifurcations}
\label{subsec-simple-regular}
\begin{figure}
\includegraphics[width=0.8\linewidth]{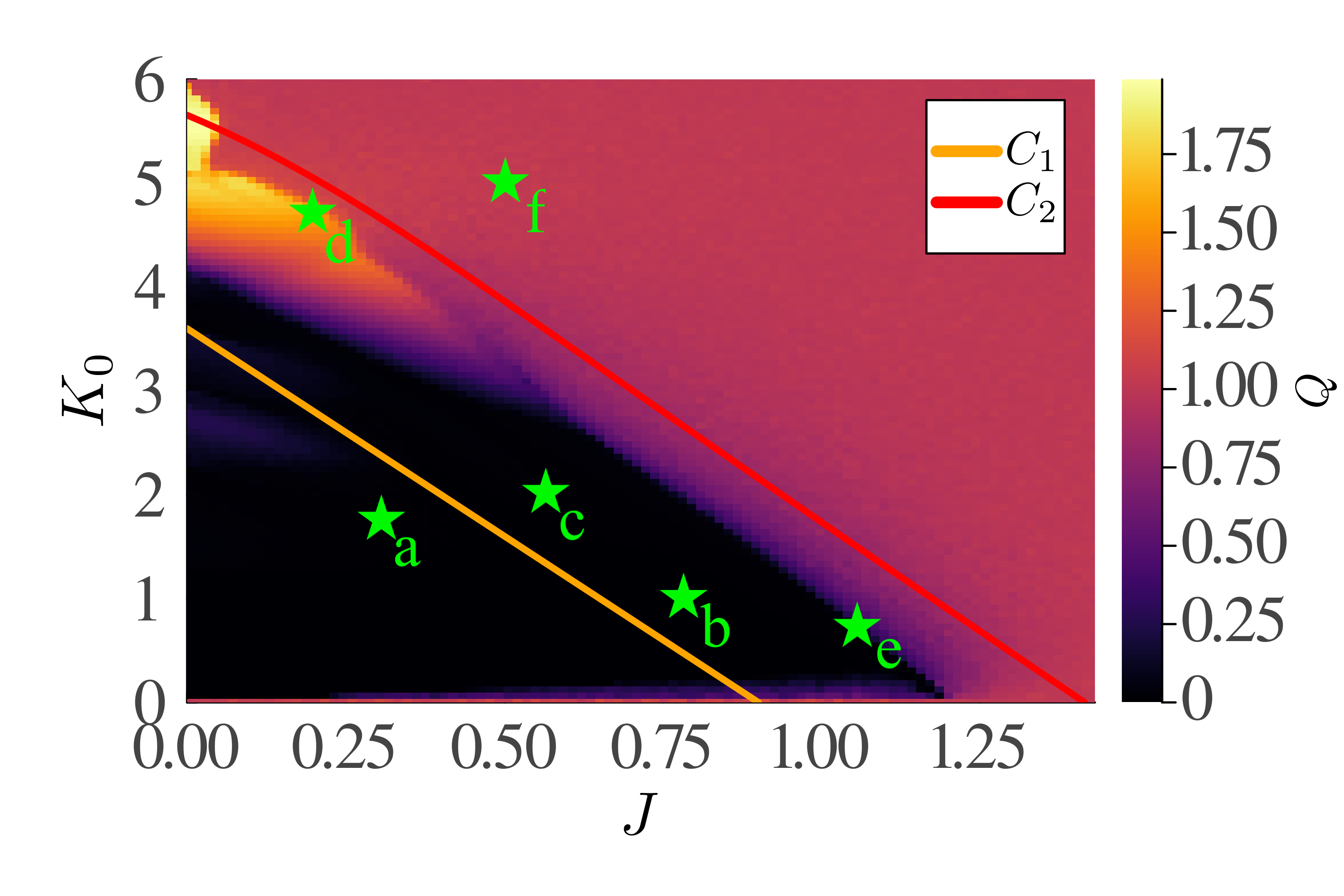}
\caption{\label{fig-C1C2} Order parameter $Q$ (Eq.\eqref{eq-Q}) for a chain of $N=100$ coupled rotors at time $t = 2000$, with the two curves $C_1$ and $C_2$. The green stars with labels (a-f) correspond to the parameter values in Fig.\ref{fig-intro}. The heatmap is generated from a $100\times 100$ grid on $J\times K_0 \in [0, 1.43]\times [0, 6]$ and averaged over $30$ random initial conditions $(p_j(0), \theta_j(0)) \in \text{Uni}[-35, 35]\times \text{Uni}[-\pi, \pi]$.}
\end{figure}

The simplest state is the stationary homogeneous state where $\theta_j = \theta^*, \forall j$. All the coupling terms $\Delta_j$ vanish and the system reduces to a single rotor. 
For $\theta^*=0$, a linear stability analysis (see Appendix \ref{appB}) gives a boundary in the parameter space where the homogeneous-zero state loses stability: 
\begin{equation}
K_0^* = -4J + 2(\gamma + 1). 
\label{eq-ckr-K0crit}
\end{equation}
We denote this curve as $C_1$, which is shown in orange on the parameter $(J, K_0)$-plane in Fig.\ref{fig-C1C2} ($\gamma=0.8$ is fixed). 
It also suggests that the instability occurs when the eigenvalue is $-1$ (cf. Appendix \ref{appB}), indicating a period-doubling bifurcation in time, simultaneously the dynamical variables alternate in space with period-$2$: $\theta_j(t) = -\theta_{j+1}(t) = -\theta_j(t+1)$ $\forall j, t$, and hence we refer it as an {\it alternating state}. 

Since an alternating state can be regarded as a steady state of the second iterated system $(\boldsymbol{p}(t+2), \boldsymbol{\theta}(t+2)) =: \boldsymbol{F}^{(2)}(\boldsymbol{p}(t), \boldsymbol{\theta}(t))$, let us denote $(\boldsymbol{p}(t+2), \boldsymbol{\theta}(t+2)) = (\boldsymbol{p}(t), \boldsymbol{\theta}(t)) =: (\boldsymbol{p}^*, \boldsymbol{\theta}^*)$ and $(\boldsymbol{p}(t+1), \boldsymbol{\theta}(t+1)) =: (-\boldsymbol{p}^*, -\boldsymbol{\theta}^*)$, where the minus signs come from 
\begin{align*}
\boldsymbol{\theta}(t+2) &= \boldsymbol{\theta}(t+1) + \boldsymbol{p}(t+2) \mod 2\pi \\
&= \boldsymbol{\theta}(t) + \boldsymbol{p}(t+1) + \boldsymbol{p}(t+2) \mod 2\pi, 
\end{align*}
which gives $\boldsymbol{p}(t+2) = -\boldsymbol{p}(t+1)$ near the bifurcation. Furthermore, we have $2\boldsymbol{\theta}^* = \boldsymbol{p}^*$. 

Now, consider the momentum equation 
\begin{equation*}
\begin{split}
p_j(t+2) &= \gamma p_j(t+1) - K_0\sin \theta_j(t+1) - J\Delta_j(t+1) \\
\Rightarrow \quad 0 &= (1 + \gamma)2\theta^* - K_0\sin \theta^* - 2J\sin (2\theta^*), 
\end{split}
\end{equation*}
where we have taken into account the spatial alternation: $\theta_j^* = -\theta_{j\pm 1}^* =: \theta^*$ and $p_j^* = -p_{j \pm 1}^* =: p^*$.
The function 
\begin{equation*}
\begin{split}
R(\theta^*) &:= (1 + \gamma)2\theta^* - K_0\sin \theta^* - 2J\sin (2\theta^*) \\
&\approx (2 + 2\gamma - K_0 - 4J)\theta^* + \frac{K_0+16J}{6}(\theta^*)^3 
\end{split}
\end{equation*}
has a unique root (which is zero) when $R'(0) > 0$ and has two additional roots when $R'(0) < 0$. The bifurcation is thus given by $R'(0) = 0$, or $2(\gamma + 1) - 4J - K_0 = 0$, for which we recover Eq.\eqref{eq-ckr-K0crit}. The approximation of $R$ suggests a pitchfork bifurcation near the origin, corresponding to a period-doubling bifurcation in the original system $\boldsymbol{F}$. Moreover, $R'''(0) = 16J + K_0 > 0$ implies that the bifurcation is supercritical. If one allows $J < 0$ and $K_0 < 0$, it becomes a subcritical bifurcation, which we will not discuss here. 

The instability of an alternating state can be determined by the eigenspectrum of the Jacobian of $\boldsymbol{F}^{(2)}$. This bifurcation curve, denoted as $C_2$, is shown in red Fig.\ref{fig-C1C2} and is generated by numerical bifurcation analysis \cite{veltz:hal-02902346}. Thus, a stable alternating state exists in the strip region in-between $C_1$ and $C_2$. 

Notice that the pure alternating state with the angle configuration $\boldsymbol{\theta}^* = (\theta^*, -\theta^*, ..., \theta^*, -\theta^*)$ can only be obtained by carefully preparing initial conditions due to multistability of the system, for example, $\boldsymbol{\theta}(0) = (1, -1, ..., 1, -1)$. With general random initial conditions, one obtains states with alternating patches as in Fig.\ref{fig-intro}(b)-(c). This will be discussed in the next section.

\subsection{Regular states inside the strip region}
\label{subsec-patch}
We observe complicated regular states inside the strip region in-between the curves $C_1$ and $C_2$, which correspond to examples (b)-(d) in Fig.\ref{fig-intro}. 
Patterns in Fig.\ref{fig-intro}(b)-(c) consist of patches with alternating feature (i.e., within a patch $\theta_j(t)=-\theta_{j+1}(t)=-\theta_j(t+1)$), so we refer them as {\it alternating-patched} (AP) states; the pattern in Fig.\ref{fig-intro}(d) consists of aligned patches (i.e., within a patch $\theta_j(t)=\theta_{j+1}(t)=-\theta_j(t+1)$), and we refer as a {\it homogeneous-patched} (HP) state. 
While those are generated from fully random initial conditions, a simpler picture can be constructed from perturbing an alternating initial state as follows. 

\begin{figure*}
\includegraphics[width=0.7\linewidth]{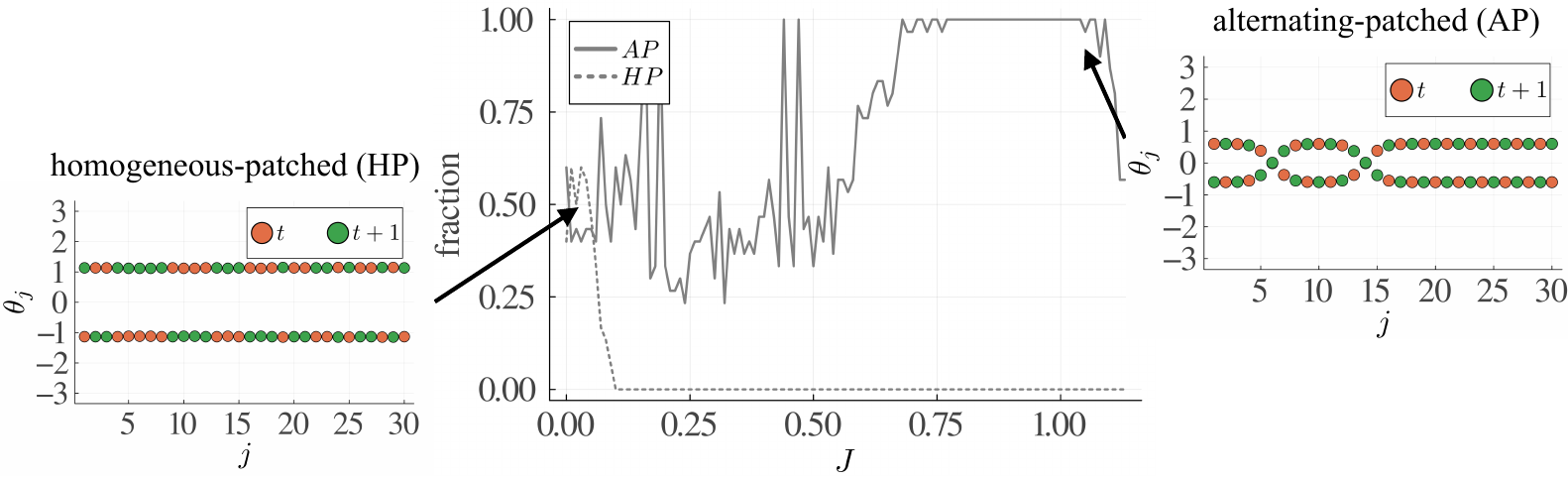}
\caption{\label{fig-patches} Fractions of alternating-patched (AP, solid) and homogeneous-patched (HP, dotted) states in varying the system parameters $(J, K_0)$ along the line $K_0(J) = K_0^*(J) + 1.0$, which is parallel to $C_1$ and inside the strip region in-between curves $C_1$ and $C_2$. Insets illustrate the two patched states: $(J, K_0) = (1.1, 0.2)$ for AP, and $(J, K_0) = (0.03, 4.48)$ for HP. Numerical settings: $N=30$, $t=10000$, and $30$ initial conditions are $\boldsymbol{p}(0) = \boldsymbol{0}$, $\boldsymbol{\theta}(0) + \boldsymbol{\epsilon}$, where $\boldsymbol{\theta}(0) = (1, -1, 1, -1, ..., 1, -1)$ and $\boldsymbol{\epsilon} = (\epsilon_1, ..., \epsilon_N)$, $\epsilon_j \in \text{Uni}[-0.01, 0.01]$, $j = 1, 2, ..., N$. }
\end{figure*}

To elucidate the relative basin sizes of AP and HP states, we perturb an initially alternating state and take parameter values along a line parallel to $C_1$ within the strip region: $K_0(J) = K_0^*(J) + 1.0$ with $J \in [0, 1.15)$. 
These two kinds of patched states can be distinguished using a local quantity
\begin{equation*}
Z := \frac{1}{2N}\sum_{j=1}^N (|\theta_j - \theta_{j-1}| + |\theta_j - \theta_{j+1}|).  
\end{equation*}
For an HP state, $Z \in (0, \overline{|\theta|})$ while for an AP state, $Z \in (\overline{|\theta|}, 2\overline{|\theta|})$, where $\overline{|\theta|} = \frac{1}{N}\sum_{j=1}^N |\theta_j|$. 
The fraction of AP and HP are shown in Fig.\ref{fig-patches}, together with their typical profiles. 

We see that the fraction of AP states undergoes strong fluctuations for $J \lesssim 0.7$, and becomes dominant for $J \in (0.7, 1.0)$. Compared to AP, the fraction of HP states slowly decreases for small $J < 0.08$ and then vanishes completely. 
Notice that AP and HP states are not the only attractors; when their fractions do not sum up to unity, additional attractors emerge. 

\subsection{Order parameter and phase diagram}
\label{subsec-order-parameter}
To better understand the rich dynamics inside the strip region, and to get a full picture of spatial patterns presented in Fig.\ref{fig-intro}, we employ order parameters widely used in phase oscillator models \cite{acebron2005kuramoto, clusella2020irregular}, in addition to the Kuramoto order parameter $Z_1 = \frac{1}{N}\sum_{j=1}^N e^{i\theta_j}$, we also consider the second harmonic, the so-called Daido order parameter, $Z_2 = \frac{1}{N}\sum_{j=1}^N e^{2i\theta_j}$, to characterize an important spatial symmetry in the system.

Consider an angle configuration with perfect spatial symmetry where there are equal numbers of $\pm \theta$ ($\theta$ is a constant), then $Z_1 = \cos \theta$ and $Z_2 = \cos 2\theta$, and thus the relation $|Z_2| = 2|Z_1|^2 - 1$ holds. We therefore denote
\begin{equation}
Q:= |Z_2| - (2|Z_1|^2 - 1) \in [-1, 2]
\label{eq-Q}
\end{equation}
as the level of deviation from this symmetry: $Q\approx 0$ when the phases are nearly $\pm \theta$-balanced, which include stationary homogeneous states, alternating states and alternating-patched (AP) states; while $Q=1$ when both $|Z_1|$ and $|Z_2|$ vanish, corresponding to a chaotic regime. Any other values of $Q$ indicate other spatial patterns such as homogeneous-patched (HP) states. 
The heatmap in Fig.\ref{fig-C1C2} shows the values of $Q$ on the parameter $(J, K_0)$-plane. First, the region below the curve $C_1$ has $Q \approx 0$ corresponding to stationary homogeneous states, and the region above the curve $C_2$ shows $Q \approx 1$ representing the chaotic regime, both as expected. Inside the strip region that is close to $C_1$, homogeneous states are bifurcated into alternating states which maintain $Q\approx 0$, and the heatmap provides additional information that, to a large extent, alternating (or AP) states are dominant (i.e., physically observable) in the strip region. But when we approach $C_2$, $Q \to 1$, indicating that the onset of chaos is earlier than loss of stability of the alternating states. In other words, chaos coexists with many regular states. 
Moreover, the period-doubling cascade of the zero state shown in the single-rotor level is suppressed \cite{bennett1990stability}, resulting in a sudden jump to chaos.

We note that the boundary of $Q=1$ in Fig.\ref{fig-C1C2} is consistent with the onset of chaos determined by the largest Lyapunov exponent in \cite{russomanno2023spatiotemporally}.
The very high value of $Q (\approx 1.75)$ in the top-left part of the strip region coincides with the relatively high fraction of HP states in Fig.\ref{fig-patches}, and this region shrinks rapidly as $J$ increases along the strip. 
In fact, $\max Q = 2$ is attained if and only if $|Z_2| = \max |Z_2| = 1$ and $|Z_1| = \min |Z_1| = 0$, which occurs under perfect symmetry with $\theta = \pm \frac{\pi}{2}$: $Z_1 = \cos \left(\pm \frac{\pi}{2}\right) = 0$ and $Z_2 = \cos (\pm \pi) = \pm 1$. Under random initial conditions, however, the final state is not perfectly symmetric. A value of $Q\approx 1.75$ can be achieved when $|Z_1| \lesssim 0.35$ and $|Z_2| \gtrsim 0.75$, with patches aligning around $\theta = \pm \frac{\pi}{2}$ in this particular parameter region. 

The parameter values used to generate patterns in Fig.\ref{fig-intro} are labeled as green stars in the heatmap: The stationary pattern in Fig.\ref{fig-intro}(a) occurs below the period-doubling bifurcation curve $C_1$ and thus of temporal period-$1$; the alternating-patched patterns in Fig.\ref{fig-intro}(b)-(c) occur inside the strip region where the order parameter $Q=0$, highlighting the alternating feature around zero momenta; Fig.\ref{fig-intro}(d) shows a mix of homogeneous and chaotic patches, as the parameter value is close to the chaos border but with a high $Q$ value; Fig.\ref{fig-intro}(e) and (f) both show chaotic patterns but (f) is apparently more spread out than the pattern in (e) -- the phase diagram concludes that patterns like (e) coexist with non-chaotic states.

Despite complex multistability in the coupled system, the bifurcations of simple states in Sec.\ref{subsec-simple-regular} can still capture the critical transitions observed in the order parameter $Q$. 
In the next section, we study an intermittent phenomenon near the onset of chaos.

\subsection{Spatiotemporal intermittency near chaos}
\label{subsec-intermittency}
As the transition to chaos is approached (purple boundary in the heatmap Fig.\ref{fig-C1C2}), a small perturbation is highly likely to trigger chaotic behavior, and the system experiences prolonged chaotic transients. 

To illustrate a long transient of an alternating-patched state, we introduce a coarse-grained {\it spin} variable according to the sign change of the momentum in space and time: 
\begin{equation*}
s_j(t) := \text{sign}\left(p_j(t)\cdot (-1)^{j+t}\right) \in \{-1, 0, 1 \}. 
\end{equation*}
When the momentum $p_j$ alternates between two values $\pm p$ both in time and space, the spin remains invariant. On the other hand, when the momentum changes around a non-zero value (e.g., $2n\pi$ with $n \neq 0$) the spin alternates the sign. 

\begin{figure*}
\includegraphics[width=0.65\linewidth]{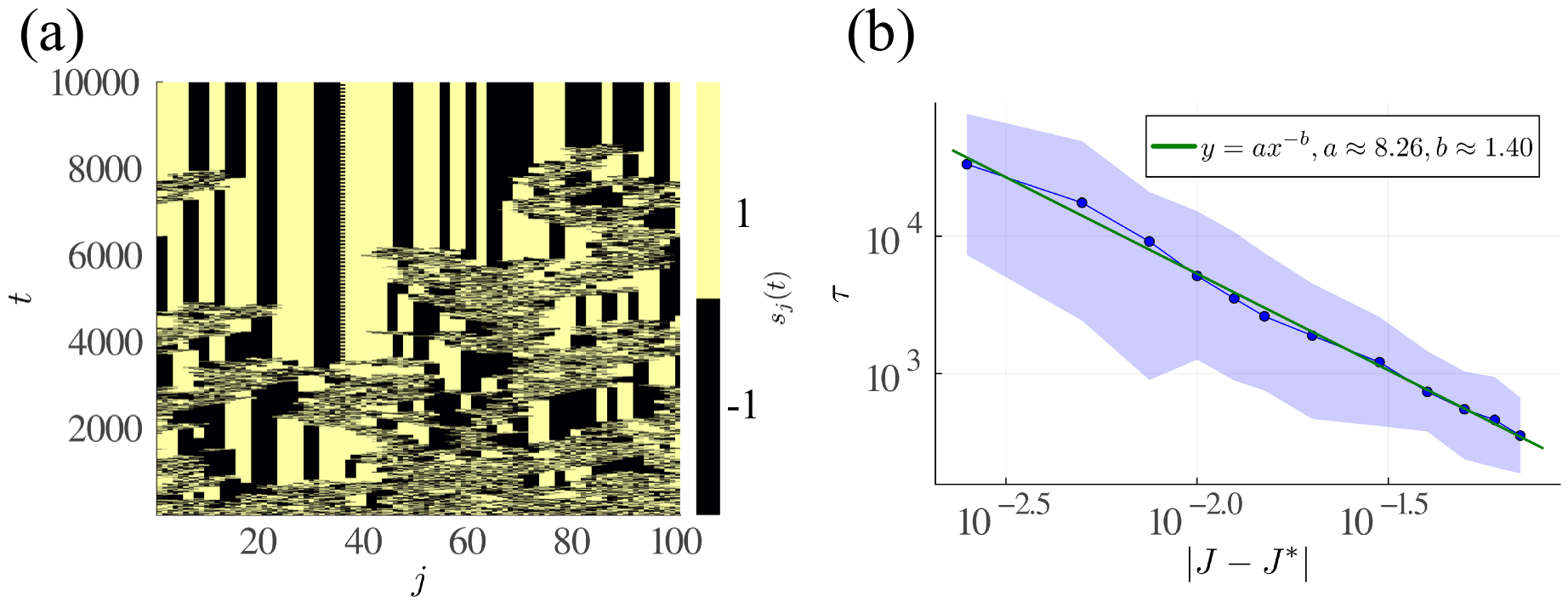}
\caption{\label{fig-spin-chaos} (a) Typical  spatiotemporal intermittent pattern of the spin $s_j(t)$, and (b) transient time $\tau$ as a function of the distance of $J$ to the critical coupling strength (onset of chaos) $J^*$ in a log-log scale. Here $\gamma=0.8$, $K_0=3.3$, $N = 100$, and the onset of chaos is estimated at $J^* = 0.5$. \cite{russomanno2023spatiotemporally} The blue curve (with data points) represents the average of $100$ trajectories, each initialized randomly $(p_j(0), \theta_j(0)) \in \text{Uni}[-35, 35]\times \text{Uni}[-\pi, \pi]$ and iterated until a steady state is reached; the blue band illustrates fluctuations. The green line is a power-law fit $y = ax^{-b}$ with $a$ and $b$ indicated in the legend.}
\end{figure*}

In Fig.\ref{fig-spin-chaos}(a), we see random patterns persist for a long time before reaching a regular state. The regular state consists of multiple alternating patches and a stationary rotor (showing in alternating colors in time). 
The transient time $\tau$ diverges as a power law near the onset of chaos $J^*$: $\tau \propto |J - J^*|^{-b}$, illustrated in Fig.\ref{fig-spin-chaos}(b). 
The exponent $b \approx 1.4$ (for $N=100$) is referred as the critical exponent of the chaotic transient \cite{grebogi1986critical, grebogi1987critical}. This behavior belongs to a class of defect turbulence with type-I super-transient at the onset of the bifurcation \cite{crutchfield1988attractors, lai2011transient}. 
In a spatially extended system one can also refer to a {\it percolation threshold} \cite{broadbent1957percolation} as the minimum concentration at which an infinite cluster spans the whole space \cite{bagnoli1999synchronization}. 

Furthermore, the average length of coherent domains (or patches) in an alternating state decreases as the system parameters approach the onset of chaos. Beyond the transition point, the system becomes fully chaotic, and the coherent domain length reduces to $1$ (i.e., a single element). In the thermodynamic limit ($N \to \infty$), this implies that the relative domain length (i.e., $1/N$) vanishes at criticality.

\section{Conclusion} 
\label{sec-conclu}
In this paper, we studied complex dynamics of a single dissipative kicked rotor and its coupled system. 
For the single map, multistability arises through multiple bifurcations, where, for the momentum variable, the principal fold bifurcating points form a cone-like boundary that restrict possible momentum values, while the principal period-doubling bifurcating points form a parabola-like boundary that separates regular and chaotic attractors when the nonlinearity $K_0$ is large. Between these two boundaries, additional branches emerge, starting with fold bifurcations and proceeding through period-doubling cascades; for small to intermediate $K_0$, these cascades terminate before developing into chaos. Only the principal branch that bifurcated from the zero fixed point continues into a chaotic attractor, however, this chaotic attractor remains bounded in momentum and can coexist with two symmetric regular branches, though the basins of these regular branches are significantly smaller than that of the chaotic attractor. 
In the large $K_0$ regime, we observe a chaotic attractor with the momentum distribution exhibiting remnants of the period-$2$ orbit bifurcated from the zero fixed point. 

For the coupled system, we provided a more microscopic picture of the dynamics compared to existing literature. The multistability of the single rotor is thus integrated into a more intricate version, whereas the local spatial patterns can still be understood via elementary solutions. We determined the stability regions of the alternating states via numerical bifurcation analysis. To address general random initial conditions, a combination of Kuramoto and Daido order parameters is employed to quantify patterns with a spatial symmetry. The rich dynamics seen from this macroscopic quantity are bounded by the bifurcation curves of the homogeneous-zero and alternating state. Additionally, this quantity indicates coexistence of regular and chaotic states near the transition to complete chaos. 

Many interesting further questions arise from our study. For example, while the sign of $K_0$ is irrelevant in the single-rotor model due to symmetry, the interplay between the signs of $K_0$ and the coupling $J$ appears to be important in the coupled system. As discussed in Sec.\ref{subsec-simple-regular}, the bifurcation of the homogeneous-zero state is supercritical when both $K_0$ and $J$ are positive, and subcritical when both are negative. Investigating transitions between these two scenarios when they have different signs would provide deeper understanding of the dynamics. 
Another open problem is to understand additional bifurcations in the region between the curves $C_1$ and $C_2$, revealing intricate structures shown in the phase diagram of the order parameter $Q$. 
Furthermore, while the basins of attraction in the single-rotor model are straightforward to visualize, analyzing them in the coupled system is significantly challenging. In this paper, we have characterized numerically the relative basin sizes of two coexisting patched states, but a full picture is still missing. For instance, it remains unclear, near the curve $C_2$, how and under what conditions the chaotic attractor dominates the whole phase space. These problems will be studied elsewhere.

\section{Acknowledgments}
The author would like to thank Matthias Wolfrum and Yuzuru Sato for useful discussions and insightful advice. 
We would also like to thank the two anonymous referees for their suggestions, which helped improve the quality and clarity of this manuscript.

\section{Author Declarations}
The author has no conflicts to disclose. 
The author confirms the sole responsibility for the conception of the study, presented results and manuscript preparation. 

\section{Data Availability}
The data that support the findings of this study are available within the article.

\appendix
\onecolumngrid
\section{\label{appA2} Probability measures on the single rotor attractors in the chaotic regime} 
\begin{figure}[H]
\includegraphics[width=0.99\linewidth]{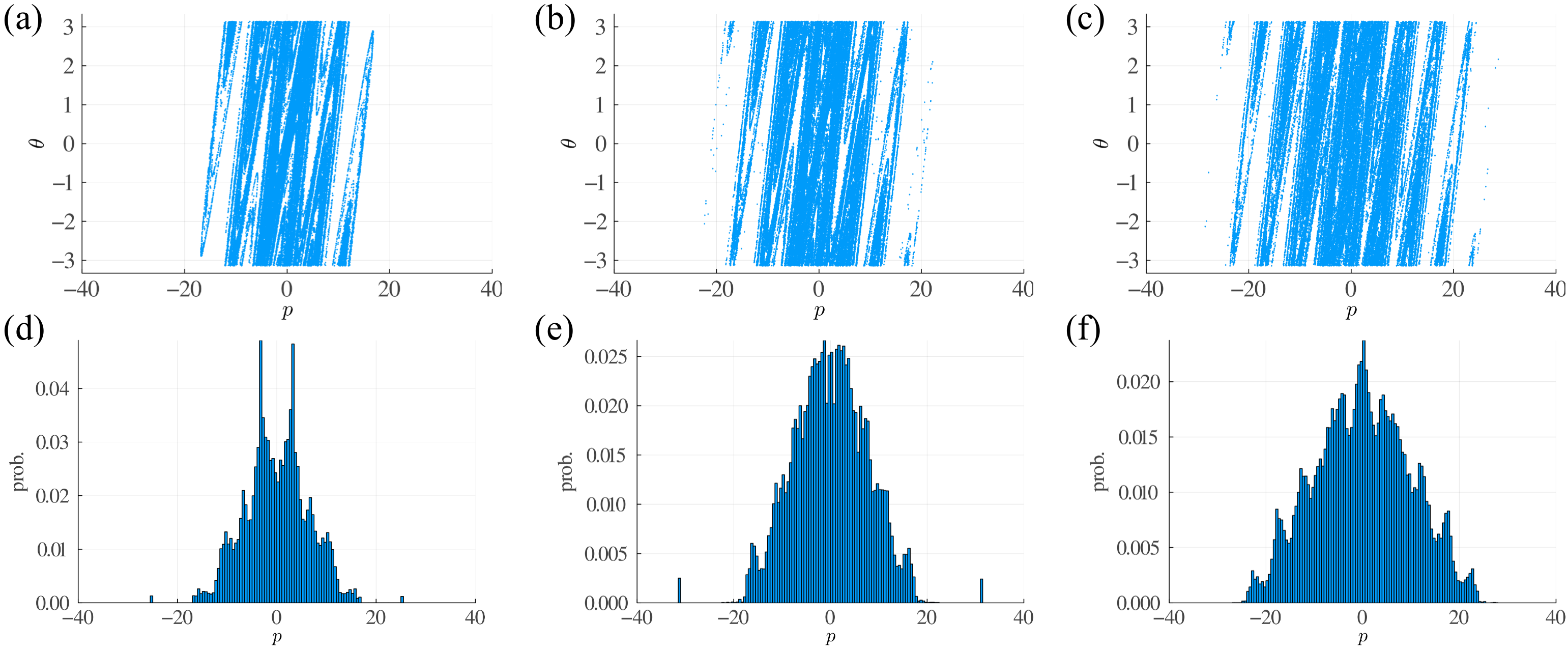}
\caption{\label{fig-chaos} Chaotic attractors (a-c) and the corresponding distributions of $p$ (d-f) for the Zaslavsky map with $\gamma = 0.8$ and $K_0 = 5.98$ (column 1), $6.6$ (column 2) and $8$ (column 3). Each attractor is generated from an arbitrary trajectory for $80000$ iterations; each histogram is generated from $50000$ trajectories starting randomly in $(p(0), \theta(0)) \in \text{Uni}[-35, 35]\times \text{Uni}[-\pi, \pi]$ for $10000$ iterations.}
\end{figure}
Fig.\ref{fig-chaos} illustrates, for three different $K_0$, the chaotic attractors and the corresponding probability distributions of the rotor momentum $p$. 
At the onset of chaos ($K_0 \approx 5.98$), a pair of distinguishable peaks near the center of the $p$-distribution is seen as remnants of the bifurcated main $0$-resonance branch. By $K_0 = 6.6$ these peaks are no longer visible, but the  comparable probabilities of the regular branches at $p = 5\cdot (2\pi) \approx \pm 31 $ are clearly visible (cf. basins in Fig.\ref{fig-basins}(c)). At $K_0 = 8$ the chaotic attractor extends further in $p$ and the $p$-distribution develops fractal-like spikes.

\section{\label{appA} Bifurcations of the single rotor in $K_0$ for other values of $\gamma$}
\begin{figure}[H]
\centering
\includegraphics[width=0.65\linewidth]{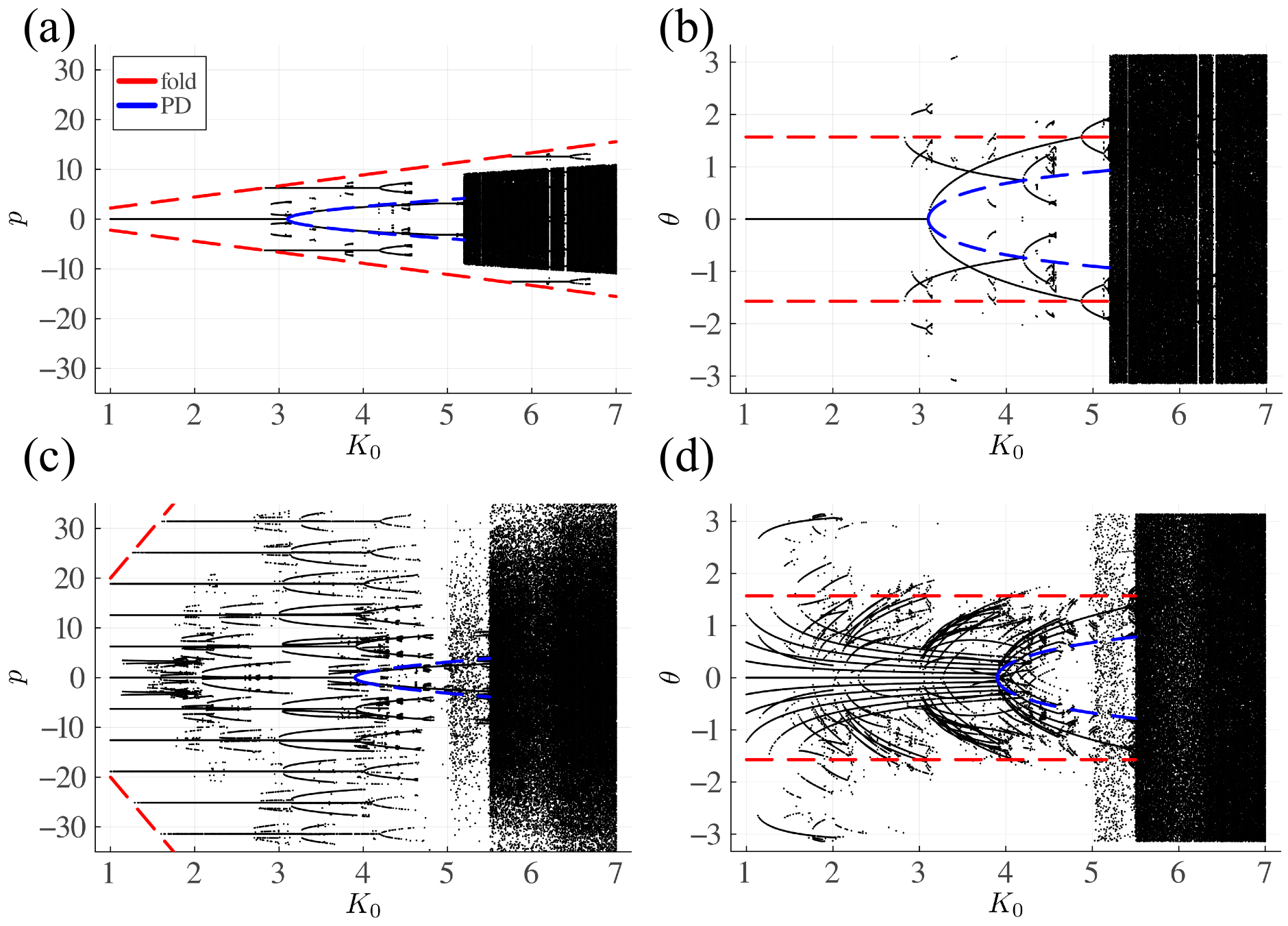} 
\caption{\label{fig-bifur-other-gammas} Bifurcations in the Zaslavsky map with $\gamma = 0.55$ (a-b) and $0.95$ (c-d). The dashed curves connect a cascade of bifurcation points: fold (in red) and period-doubling (PD in short, in blue) bifurcations for fixed points $p^*=2n\pi$, $n\in \mathbb{Z}$.}
\end{figure}

The dissipation coefficient $\gamma$ affects the momentum range, as clearly illustrated in the bifurcation diagrams in Fig.\ref{fig-bifur-other-gammas}. The cascades of bifurcating branches discussed in Sec.\ref{subsec-sr-bifur} are present for any $\gamma \in (0, 1)$.

\section{\label{appB} Linear stability analysis for the coupled system}
The linearized equations of motion around the zero state $p_j = 0$, $\theta_j = 0$ $\forall j = 1, 2, ..., N$ read 
\begin{equation*}
\begin{split}
& p_j(t+1) = \gamma p_j(t) - J[2\theta_j(t) - \theta_{j-1}(t) - \theta_{j+1}(t)] - K_0\theta_j(t), \\
& \theta_j(t+1) = \theta_j(t) + p_j(t+1) \quad (\text{mod } 2\pi). 
\end{split}
\end{equation*}
Applying a Fourier transform $p_j(t) = \sum_w P_w(t) e^{iwj}$, $\theta_j(t) = \sum_w\Theta_w(t) e^{iwj}$, $w = \frac{2\pi l}{N}$, $l = 0, 1, ..., N-1$ (for periodic boundary conditions) gives, for each pair of Fourier variables $(P_w, \Theta_w$), 
\begin{equation*}
\begin{pmatrix} 
P_w(t+1) \\
\Theta_w(t+1)
\end{pmatrix}
= \begin{pmatrix}
\gamma & -[2J(1 - \cos w) + K_0] \\
\gamma & 1 - [2J(1 - \cos w) + K_0] 
\end{pmatrix}
\begin{pmatrix}
P_w(t) \\
\Theta_w(t)
\end{pmatrix}, 
\end{equation*}
whose characteristic equation is 
\begin{equation}
\lambda^2 - [\gamma + 1 - 2J(1 - \cos w) - K_0]\lambda + \gamma = 0,
\label{eq-linear-chara}
\end{equation}
and its solutions are given by  
\begin{equation*}
\lambda_w^{\pm} = \frac{1}{2}\left[ \gamma + 1 - 2J(1 - \cos w) - K_0 \pm \sqrt{[\gamma + 1 - 2J(1 - \cos w) - K_0]^2 - 4\gamma}\right].
\end{equation*}
We acknowledge that the above derivation was previously carried out in \cite{russomanno2023spatiotemporally}. Here, we extend the analysis by proving that the critical case corresponds to the eigenvalue crossing the unit circle at $-1$, with the associated Fourier mode being $l=\frac{N}{2}$.

\begin{figure}[H]
\centering
\includegraphics[width=0.55\linewidth]{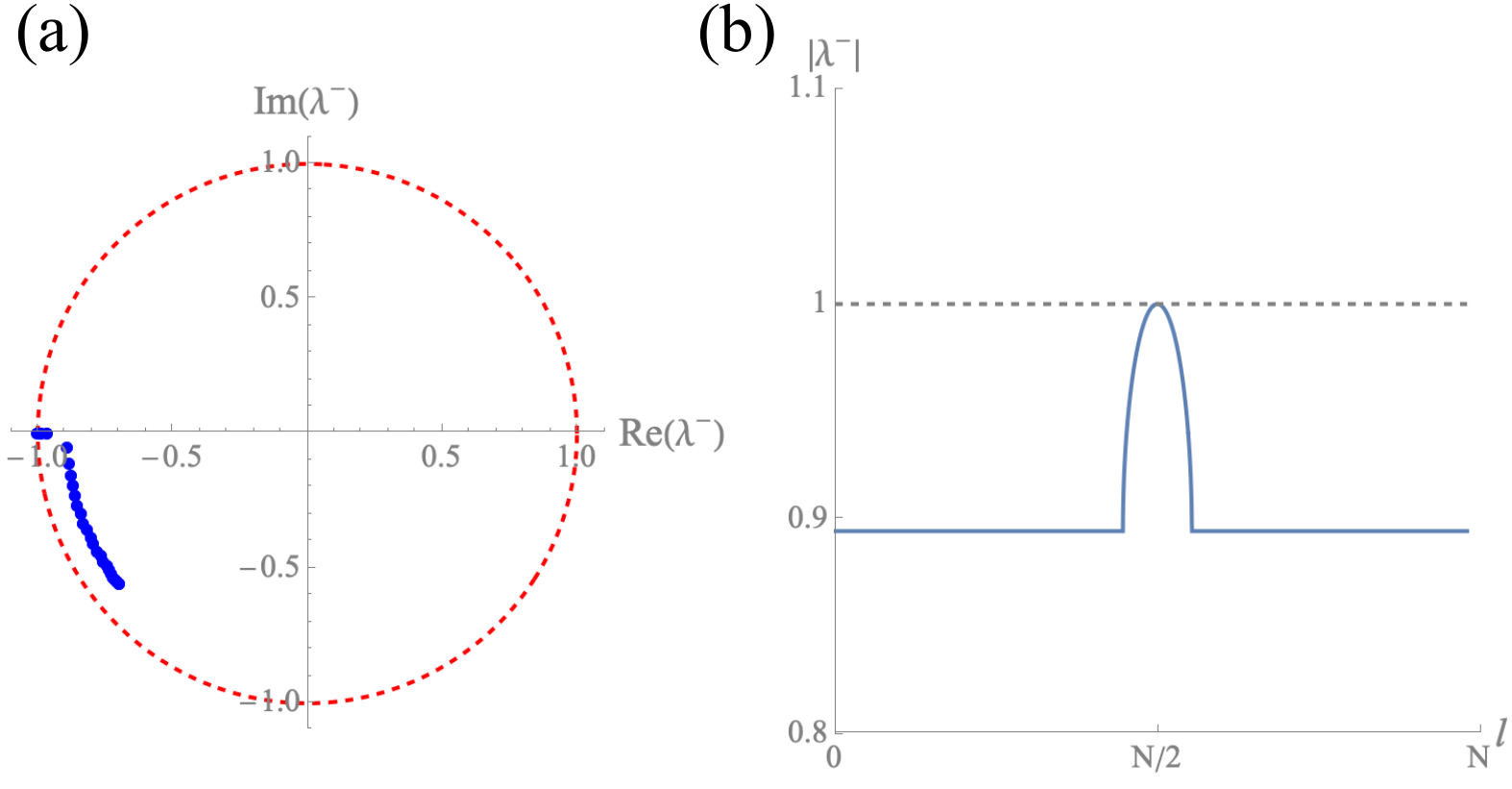}
\caption{\label{fig-evalues} (a) Eigenvalues $\lambda^-_w$ and (b) their magnitudes at the critical transition: $\gamma=0.8$, $K_0 = 3.2$, $J=0.1$.}
\end{figure}

The homogeneous-zero solution becomes unstable when there is an eigenvalue with modulus larger than one. An example is illustrated in Fig.\ref{fig-evalues}. 

We now prove that this critical transition occurs when $\lambda_{w=\pi}^- = -1$. 
 
It is clear that for $\lambda_w^+$, the maximum is attained when $\cos w = 1$, or $w = 0$: 
\begin{equation*}
\max_w \lambda_w^+ = \lambda_0^+ = \frac{1}{2}\left[ \gamma + 1 - K_0 + \sqrt{[\gamma + 1 - K_0]^2 - 4\gamma}\right].
\end{equation*}
When the term in the square-root is negative, i.e., $(\sqrt{\gamma} - 1)^2 < K_0 < (\sqrt{\gamma} + 1)^2$, we have 
\begin{equation*}
|\lambda_0^+| = \frac{1}{2}\sqrt{(\gamma + 1 - K_0)^2 - (\gamma + 1 - K_0)^2 + 4\gamma} = \sqrt{\gamma} < 1. 
\end{equation*}
Otherwise, we have $|\lambda_0^+| = \frac{1}{2}|(\gamma + 1 - K_0) + \sqrt{(\gamma + 1 - K_0)^2 - 4\gamma}|$ as a decreasing function in $K_0 (>0)$, and thus $|\lambda_0^+| < |\lambda_0^+|_{K_0 = 0} = \frac{1}{2}|(\gamma + 1) + \sqrt{(\gamma + 1)^2 - 4\gamma}| = 1$. In summary, $|\max_w \lambda_w^+| < 1$ for all parameter values. 

For $\lambda_w^-$, when the term in the square-root is negative, we have again 
\begin{equation*}
|\lambda_w^-| = \frac{1}{2}\left[ (\gamma + 1 - 2J(1 - \cos w) - K_0)^2 - [\gamma + 1 - 2J(1 - \cos w) - K_0]^2 + 4\gamma\right] = \sqrt{\gamma} < 1. 
\end{equation*}
Otherwise, let us denote $A := \gamma + 1 - 2J(1 - \cos w) - K_0$. The condition of $\lambda_w^+ \in \mathbb{R}$ can be written as $A^2 - 4\gamma \geq 0$, or equivalently, $A \leq -2\sqrt{\gamma}$ or $A \geq 2\sqrt{\gamma}$. Furthermore, $\lambda_w^- := f(A)$ becomes 
\begin{equation*}
f(A) = \frac{1}{2}(A - \sqrt{A^2 - 4\gamma}). 
\end{equation*} 
From $f'(A) = \frac{1}{2} - \frac{A}{2\sqrt{A^2 - 4\gamma}}$ we have $f'(A) < 0$ when $A > 2\sqrt{\gamma}$, so $\max f = f(A = 2\sqrt{\gamma}) = \sqrt{\gamma} < 1$; on the other hand, when $A < -2\sqrt{\gamma}$, $f'(A) > 0$ and $\max_A f(A) = f(A = -2\sqrt{\gamma}) = -\sqrt{\gamma}$. Moreover, $\lim_{A \to -\infty} f'(A) = -\infty$. Therefore, for $A \leq -2\sqrt{\gamma}$ we have $f(A) \in (-\infty, -\sqrt{\gamma}]$. The maximum of $|f(A)|$ over all $w$ is attained when $A = A(w)$ is minimum, i.e., when $\cos w = -1$, or $w = \pi$. 
The crossing of the unit circle thus happens at $\lambda_w^- = -1 = \frac{1}{2}(A - \sqrt{A^2 - 4\gamma})$, or $A(\pi) = -(1 + \gamma)$. Substituting in the definition of $A$ gives 
\begin{equation*}
K_0^* = -4J + 2(\gamma + 1).
\end{equation*}
One can also simply plug $\lambda = -1$ and $w = \pi$ into Eq.\eqref{eq-linear-chara}. 
Note that the critical curve given by eq.(8) in \cite{russomanno2023spatiotemporally} is not precise, where the author substituted $w=\pi$ but incorrectly letting the square-root part of the eigenvalue $\lambda_{w=\pi}$ vanish, which does not lead to the critical case $|\lambda_{w=\pi}|=1$, but rather $|\lambda_{w=\pi}| = \sqrt{\gamma} < 1$. This discrepancy becomes more apparent for smaller $\gamma$.

We conclude that $w = \pi$ represents the most unstable mode, which corresponds to $l = \frac{N}{2}$ in the Fourier mode $w := \frac{2\pi l}{N}$. It implies that the dynamical variables alternate in space with period-$2$; the eigenvalue crossing the unit circle at $-1$ indicates a period-$2$ bifurcation in time.

\section*{References}
\twocolumngrid
\bibliography{Ref}

\end{document}